\def\ra{\rangle}
\def\la{\langle}
\def\be{\begin{equation}}
	\def\ee{\end{equation}}
\def\ba{\begin{array}}
	\def\ea{\end{array}}

\makeatletter

\newcommand{\Rmnum}[1]{\expandafter\@slowromancap\romannumeral #1@}
\makeatother

\documentclass[11pt,a4paper]{article}
\usepackage[T1]{fontenc}
\usepackage[utf8]{inputenc}
\usepackage{authblk}
\usepackage{subfigure}
\usepackage[colorlinks=True,linkcolor=blue,anchorcolor=blue,citecolor=blue,urlcolor=blue,CJKbookmarks=true]{hyperref}
\usepackage{amsfonts}
\usepackage{flafter}
\usepackage{amssymb,graphics,graphicx,subfigure,caption2,rotating}
\usepackage[numbers,sort&compress]{natbib}
\usepackage{amsmath}
\usepackage{indentfirst}
\usepackage{amsmath, latexsym}
\usepackage[amsmath,thmmarks]{ntheorem}
\usepackage[top=3.5cm]{geometry}
\input amssym.def

\begin{document}
\date{}
	\title{\textbf{A note on the lower bounds of genuine multipartite entanglement concurrence}}
	\author[1]{Ming Li \footnote{E-mail: liming@upc.edu.cn.}}
	\author[1]{Yaru Dong}
	\author[1]{Ruiqi Zhang}
	\author[2]{Xuena Zhu}
	\author[1]{Shuqian Shen}
	\author[1]{Lei Li}
	\author[3,4]{Shao-Ming Fei}
	\affil[1]{College of the Science, China University of	Petroleum, 266580 Qingdao, China}
	\affil[2]{School of Mathematics and Statistics Science, Ludong University, Yantai 264025, China}
	\affil[3]{School of Mathematical Sciences, Capital Normal University, Beijing 100048, China}
	\affil[4]{Max-Planck-Institute for Mathematics in the Sciences, 04103 Leipzig, Germany}
		{
	}\maketitle

\begin{abstract}
Quantum entanglement plays a pivotal role in quantum information processing. Quantifying quantum entanglement is a challenging and essential research area within the field. This manuscript explores the relationships between bipartite entanglement concurrence, multipartite entanglement concurrence, and genuine multipartite entanglement (GME) concurrence. We derive lower bounds on GME concurrence from these relationships, demonstrating their superiority over existing results through rigorous proofs and numerical examples. Additionally, we investigate the connections between GME concurrence and other entanglement measures, such as tangle and global negativity, in multipartite quantum systems.
	
\end{abstract}

\section{Introduction}
Quantum entanglement is a foundational resource in quantum information processing, enabling advancements in tasks like quantum teleportation, quantum dense coding, quantum computation, and quantum state preparation \cite{Bouwmeester, Riedmatten, Mattle, nielsen, di, Resch}. Therefore, accurately characterizing and quantifying entanglement is of paramount importance.

In recent years, various measures have been proposed to quantify bi- and multipartite entanglement \cite{Vedral, chen kai, Horodecki, Szalay, Yu C.S}. One prominent measure is concurrence, initially defined for two-qubit systems \cite{W.K.Wootters, G. Gour, Hill}. Wootters obtained an elegant formula to compute the concurrence for arbitrary mixed states of two qubits \cite{Wootters}.  Mintert then generalized the definition further to multipartite systems, introducing multipartite entanglement concurrence \cite{Mintert}. Additionally, Ma introduced GME concurrence, a variant derived from the classic concurrence, specifically to quantify genuine multipartite entanglement \cite{ma1}. These measures serve to assess the extent of entanglement present in quantum systems.

Given the computational challenges associated with determining entanglement measures, particularly their NP-hard nature \cite{nphard}, establishing effective lower bounds becomes critical. Recent research efforts have focused on developing rigorous lower bounds for GME concurrence using various methodologies, highlighting their significance in assessing multipartite entanglement accurately and efficiently \cite{Li Ming1,Li Ming2,W. Xu,J. Wang}.

In this manuscript, we explore the interconnections between bipartite entanglement concurrence, multipartite entanglement concurrence, and genuine multipartite entanglement (GME) concurrence. Specifically, we investigate the relationship between the lower bounds of GME concurrence and other entanglement measures in multipartite quantum systems. Our approach involves deriving and evaluating these lower bounds, showcasing their effectiveness compared to existing results.

The structure of this paper unfolds as follows:
Section II introduces essential definitions and concepts.
Section III examines the relationship between bipartite concurrences and GME concurrences, leading directly to the derivation of lower bounds for GME concurrence. We also assess the practical efficacy of these bounds through comparisons with established findings.
Section IV delves into the relationship between multipartite entanglement concurrences and GME concurrences.
Section V explores the connections between GME concurrence and other entanglement metrics such as tangle and global negativity, building upon the insights gained in previous sections.
Finally, we draw conclusions in the last section, summarizing the implications of our findings and suggesting directions for future research.

\section{Preliminaries}

We first introduce some definitions and notations. Let $H_i^d$, $i=1,2,\cdots,N$, denote the $d$-dimensional Hilbert spaces.
An $N$-partite state $\rho  \in H_1\otimes H_2 \otimes\cdots\otimes H_N$ can be expressed as $\rho  = \sum {{p_\alpha }} \left| {{\psi _\alpha }} \right\rangle \left\langle {{\psi_\alpha }} \right|$, where $0<p_\alpha\leq 1$, $\sum {{p_\alpha }}  = 1$, $\left| {{\psi _\alpha }} \right\rangle \in H_1^d \otimes H_2^d\otimes\cdots\otimes H_N$ are normalized pure states.
If all $\left| {{\psi _\alpha }} \right\rangle$ are bi-separable, namely, $\left| {{\psi _\alpha }} \right\rangle = \left| {\varphi _\alpha ^x} \right\rangle  \otimes \left| {\varphi _\alpha ^{\overline{x}}} \right\rangle $, where $x$ is a certain nonempty subset of $\{1,2,\cdots,N\}$, $\overline{x}$ is the complement of $x$, 
then $\rho $ is said to be bipartite separable. Otherwise, $\rho $ is called genuine $N$-partite entangled.

For a bipartite pure state $\left| \varphi  \right\rangle \in {H_{AB}}={H_A} \otimes {H_B}$, where
$H_A$ ($H_B$) denotes the $m$ ($n$)-dimensional vector space associated with the subsystem $A$ ($B$)
such that $m \le n$, the concurrence \cite{Hill,C.S. Yu} is defined by
\begin{align*}
C_{A|B}\left( {\left| \varphi  \right\rangle } \right) = \sqrt {2\left(1 -{\rm
		tr}\rho_{A}^2 \right)},
\end{align*}
with the reduced matrix ${\rho_A}$ obtained by tracing over the subsystem $B$ and $\rm tr(\cdot)$ denotes the tracing of the density matrix.
The concurrence is then extended to mixed states $\rho\in {H_{AB}}$ by the convex roof \cite{C.J. Zhang}:
\begin{equation}\label{bicon}
	C_{A|B}\left( \rho  \right) \equiv \mathop {\min }\limits_{\left\{ {{p_i},\left| {{\varphi _i}} \right\rangle } \right\}} \sum\limits_i {{p_i}C_{A|B}\left({\left| {{\varphi_i}} \right\rangle} \right)},
\end{equation}
where the minimum is taken over all possible ensemble decompositions of $\rho  = \sum\limits_i {{p_i}\left| {{\varphi _i}} \right\rangle } \left\langle {{\varphi _i}} \right|$, ${p_i} \ge 0$ and $\sum\limits_i {{p_i}}  = 1$.

The definition of concurrence is then generalized to multipartite
case. Let $|\psi\ra\in H_{1}\otimes H_{2}\otimes\cdots\otimes H_{N}$, $dim H_{i}=d$, $i=1,\cdots,N$ be a pure $N$-partite quantum state, the
concurrence of $|\psi\ra$ is defined by \cite{multicon}
\begin{eqnarray}\label{xxxx}
	C_{N}(|\psi\ra\la\psi|)=2^{1-\frac{N}{2}}\sqrt{(2^{N}-2)-\sum_{\alpha}{\rm
			tr}\rho_{\alpha}^{2}},
\end{eqnarray}
where $\alpha$ labels all different reduced density matrices. When $N=2$, (\ref{xxxx}) reduces to (\ref{bicon}).

For a mixed multipartite quantum state,
$\rho=\sum_{i}p_{i}|\psi_{i}\ra\la\psi_{i}|$ in $H_{1}\otimes H_{2}\otimes\cdots\otimes H_{N}$,
the corresponding concurrence (\ref{xxxx}) is then given by the
convex roof:
\begin{eqnarray}\label{def}
	C_{N}(\rho)=\min_{\{p_{i},|\psi_{i}\ra\}}\sum_{i}p_{i}C_{N}(|\psi_{i}\ra\la\psi_{i}|).
\end{eqnarray}

For an $N$-partite pure state $|\psi\ra\in H_1\otimes H_2 \otimes\cdots\otimes H_N$, the GME concurrence $C_{GME}(|\psi\ra)$ is defined by
\begin{eqnarray*}
	C_{GME}(|\psi\ra)=\sqrt{\min_x\{1-{\rm
			tr}\rho_x^2\}},
\end{eqnarray*}
where $\rho_x$ is the reduced matrix with respect to subset $x$, namely, the minimum is taking over all the reduced density matrices.
This GME concurrence is proved to be a well defined measure for $N$-partite quantum systems \cite{ma1,ma2}.
For mixed states $\rho\in H_1\otimes H_2 \otimes\cdots\otimes H_N$, the GME concurrence is then defined by the convex roof
\begin{eqnarray}
	C_{GME}(\rho)=\min_{\{p_{\alpha},|\psi_{\alpha}{\ra\}}}\sum_{\alpha}p_{\alpha}C_{GME}(|\psi_{\alpha}\ra),
\end{eqnarray}
where minimum is taken over all pure ensemble decompositions of $\rho$.
Due to the optimal minimization involved in the estimation of $C_{GME}(\rho)$, it is generally quite hard to compute
the GME concurrence. Instead of exact expressions of $C_{GME}(\rho)$ for pure states or some density matrices with special form, it is usually possible to give bounds on the GME concurrence.

\section{A relationship between bipartite concurrences and GME concurrences}
In this section, we first analyze and discuss the relationship between the GME concurrence and the bipartite concurrences in tripartite quantum systems, and then generalize this result to multipartite systems. Computing $C_{GME}(\rho)$ is difficult because of the optimality minimization required in estimating $C_{GME}(\rho)$. However, we can give the following lower bound for the $C_{GME}(\rho)$.

\textbf{Theorem 1:}
\emph{Let $\rho  \in {H_{123}} = H_1^d \otimes H_2^d \otimes H_3^d$ be a tripartite qudits quantum state.
Then one has
\be\label{THM}
C_{GME}(\rho)\geq \frac{C_{1|23}(\rho)+C_{2|13}(\rho)+C_{3|12}(\rho)}{\sqrt{2}}
-2{\sqrt{\frac{d-1}{d}}},
\ee
where $C_{i|jk}(\rho)$, $i\neq j\neq k=1,2,3$, is the concurrence under the bipartition $i|jk$.}\\
\textbf{Proof.} We consider pure state first.
For $\rho=|\psi\ra\la\psi|\in H_1^d \otimes H_2^d \otimes H_3^d$, we have
\begin{align*}
	&\sqrt{d(d-1)}\sqrt{1-{\rm tr}\rho_1^2}-\sqrt{\frac{d(d-1)}{2}}\left(C_{1|23}(|\psi\ra)+C_{2|13}(|\psi\ra)+C_{3|12}(|\psi\ra)\right)-2+2d\notag\\
	=&\sqrt{d(d-1)}\sqrt{1-{\rm tr}\rho_1^2}-\sqrt{d(d-1)}\left(\sqrt{1-{\rm tr}\rho_1^2}+\sqrt{1-{\rm tr}\rho_2^2}
	+\sqrt{1-{\rm tr}\rho_3^2}\right)-2+2d\notag\\
	=&-\sqrt{d(d-1)}\left(\sqrt{1-{\rm tr}\rho_2^2}+\sqrt{1-{\rm tr}\rho_3^2}\right)-2+2d\notag\\
	\geq&0,
\end{align*}
where we have used the inequality $\sqrt{1-{\rm tr}\rho_{k}^2}\leq \sqrt{1-\frac{1}{d}}$ for $k=2$ or $3$ to obtain the inequality.
Thus we get
\be
\sqrt{1-{\rm tr}\rho_1^2}\geq\frac{C_{1|23}(|\psi\ra)+C_{2|13}(|\psi\ra)
	+C_{3|12}(|\psi\ra)}{\sqrt{2}}
-2\sqrt{\frac{d-1}{d}}.\ee
Similarly we obtain
\be\sqrt{1-{\rm tr}\rho_k^2}\geq\frac{C_{1|23}(|\psi\ra)+C_{2|13}(|\psi\ra)
	+C_{3|12}(|\psi\ra)}{\sqrt{2}}
-2\sqrt{\frac{d-1}{d}},\ee
where $k=2,3.$

Then according to the definition of GME concurrence, we derive
\be
C_{GME}(|\psi\ra)\geq \frac{C_{1|23}(|\psi\ra)+C_{2|13}(|\psi\ra)
	+C_{3|12}(|\psi\ra)}{\sqrt{2}}
-2\sqrt{\frac{d-1}{d}}.
\ee

Now consider a mixed state $\rho\in H_1^d \otimes H_2^d \otimes H_3^d$ with the optimal ensemble decomposition
$\rho=\sum_{\alpha}p_{\alpha}|\psi_{\alpha}\ra\la\psi_{\alpha}|$ such that the GME concurrence attains its minimum.
By using the convexity property of concurrence, one gets that
\begin{align*}
	&C_{GME}(\rho)=\sum_{\{p_{\alpha},|\psi_{\alpha}\ra\}}p_{\alpha}C_{GME}(|\psi_{\alpha}\ra)\\
	\geq& \sum_{\alpha}p_{\alpha}\left(\frac{C_{1|23}(|\psi_{\alpha}\ra)+C_{2|13}(|\psi_{\alpha}\ra)
		+C_{3|12}(|\psi_{\alpha}\ra)}{\sqrt{2}}
	-2\sqrt{\frac{d-1}{d}}\right) \\
	\geq&\left( \frac{C_{1|23}(\rho)+C_{2|13}(\rho)+C_{3|12}(\rho)}{\sqrt{2}}\right) 
	-2\sqrt{\frac{d-1}{d}},
\end{align*}
which ends the proof of the Theorem.
\hfill \rule{1ex}{1ex}

The above theorem gives a relation between the tripartite genuine entanglement concurrence and the bipartite entanglement concurrence. In the following we generalize the lower bound to arbitrary $N$-partite systems.

\textbf{Theorem 2:}
\emph{Let $\rho  \in {H_{12\cdots N}} = H_1^d \otimes H_2^d \otimes\cdots\otimes H_N^d$ be an $N$-partite qudits quantum state.
Then one has
\be\label{THM3}
C_{GME}(\rho)\geq \frac{1}{\sqrt{2}}\sum_{x}{C_{x|\overline{x}}(\rho)}
-\left( 2^{N-1}-2\right) {\sqrt{\frac{d-1}{d}}},
\ee
where $C_{x|\overline{x}}(\rho)$ is the concurrence under the bipartition $x|\overline{x}$, and the summation term is taken over all $2^{N-1}-1$ different partitions $($It should be noted that $x|\overline{x}$ and $\overline{x}|x$ are identical partitions, for example, $1|2\cdots N $ is the same partition as $2\cdots N|1)$, and  $x,\overline{x}\in\mathcal{P}(\mathcal{S})\setminus \left\lbrace \varnothing,\mathcal{S}\right\rbrace,x\cup\overline{x}=\mathcal{S},x\cap\overline{x}=\varnothing, \mathcal{S}=\left\lbrace 1,2,3,\cdots,N\right\rbrace$, and $$\mathcal{P}(\mathcal{S})=\left\lbrace  \varnothing,\left\lbrace 1\right\rbrace ,\cdots ,\left\lbrace N\right\rbrace ,\cdots ,\left\lbrace (N-1),N\right\rbrace ,\cdots,\left\lbrace 1,2,3,\cdots,N\right\rbrace \right\rbrace $$ denotes the power set of $\mathcal{S}$.} \\
\textbf{Proof.} Similarly, we first consider the pure state. For any state $\rho=|\psi\rangle\langle\psi|  \in {H_{12\cdots N}} = H_1^d \otimes H_2^d \otimes\cdots\otimes H_N^d$, and all $x,\overline{x}\in\mathcal{P}(\mathcal{S})\setminus \left\lbrace \varnothing,\mathcal{S}\right\rbrace$, we have \begin{align*}
	&\sqrt{d(d-1)}\sqrt{1-{\rm tr}\rho_1^2}-\sqrt{\frac{d(d-1)}{2}}\sum_{x}C_{x|\overline{x}}(|\psi\rangle)-(2^{N-1}-2)+(2^{N-1}-2)d\notag\\
	=&\sqrt{d(d-1)}\sqrt{1-{\rm tr}\rho_1^2}-\sqrt{d(d-1)}\sum_{x}\sqrt{1-{\rm tr}\rho_{x}^2}-(2^{N-1}-2)+(2^{N-1}-2)d\notag\\
	=&-\sqrt{d(d-1)}\sum_{x\ne1 }\sqrt{1-{\rm tr}\rho_{x}^2}-(2^{N-1}-2)+(2^{N-1}-2)d\notag\\
	\geq&0,
\end{align*}
where we have used the inequality $\sqrt{1-{\rm tr}\rho_{x}^2}\leq \sqrt{1-\frac{1}{d}}$ for all the different partitions $x|\overline{x}$, $x,\overline{x}\in \mathcal{P}(\mathcal{S})\setminus \left\lbrace \varnothing,\mathcal{S},\left\lbrace 1\right\rbrace \right\rbrace $ in the second equation.
Thus for all the different partitions $k|\overline{k}$, $k,\overline{k}\in \mathcal{P}(\mathcal{S})\setminus \left\lbrace \varnothing,\mathcal{S}\right\rbrace $, we get
\be 
\label{11}
\sqrt{1-{\rm tr}\rho_k^2}\geq \frac{1}{\sqrt{2}}\sum_{x}C_{x|\overline{x}}(|\psi\rangle)-(2^{N-1}-2)\sqrt{\frac{d-1}{d}}.
\ee 
Then according to the definition of GME concurrence, we obtain 
\be
C_{GME}(|\psi\rangle)\geq \frac{1}{\sqrt{2}}\sum_{x}C_{x|\overline{x}}(|\psi\rangle)-(2^{N-1}-2)\sqrt{\frac{d-1}{d}}.
\ee
We consider taking the optimal decomposition $\rho=\sum_{\alpha}p_{\alpha}|\psi_{\alpha}\ra\la\psi_{\alpha}|$ for the mixed state $\rho\in  H_1^d \otimes H_2^d \otimes\cdots\otimes H_N^d$ to minimize the GME concurrence.
According to the convexity property of concurrence, we get that
\begin{align*}
	&C_{GME}(\rho)=\sum_{\{p_{\alpha},|\psi_{\alpha}\ra\}}p_{\alpha}C_{GME}(|\psi_{\alpha}\ra)\\
	&\geq \sum_{\alpha}p_{\alpha}\left( \frac{1}{\sqrt{2}}\sum_{x}C_{x|\overline{x}}(|\psi_\alpha \rangle)-(2^{N-1}-2)\sqrt{\frac{d-1}{d}}
	\right) \\
	&\geq\frac{1}{\sqrt{2}}\sum_{x}C_{x|\overline{x}}(\rho)-(2^{N-1}-2)\sqrt{\frac{d-1}{d}},
\end{align*}
which leads to the result of the Theorem.
\hfill \rule{1ex}{1ex}\par
\textbf{Remark 1.} Notice that Theorem 2 simplifies to Theorem 1 when $N=3$. And we can show that the proposed lower bound on the GME concurrence improves some existing lower bounds \cite{Li Ming2}.
 
\textbf{Theorem 3:}
\emph{Let $\rho  \in {H_{123}} = H_1^d \otimes H_2^d \otimes H_3^d$ be a tripartite qudits quantum state.
Then we have the lower bound of GME concurrence 
\begin{align*}
	C_{GME}(\rho)\geq \frac{C_{1|23}(\rho)+C_{2|13}(\rho)+C_{3|12}(\rho)}{\sqrt{2}}
	-2{\sqrt{\frac{d-1}{d}}}
\end{align*}
 is tighter than the lower bound of GME concurrence
 \begin{align*}
 	C_{GME}(\rho)\geq \frac{1}{\sqrt{d(d-1)}}\left( \max\left\lbrace M(\rho),N(\rho)\right\rbrace -\frac{1+2d}{3}\right):=L_1(\rho)
 \end{align*}
 given in \cite{Li Ming2}, where $M(\rho)=\frac{1}{3}(||\rho^{T_1}||_{KF}+||\rho^{T_2}||_{KF}+||\rho^{T_3}||_{KF}), N(\rho)=\frac{1}{3}(||R_{1|23(\rho)}||_{KF}+||R_{2|13(\rho)}||_{KF}+||R_{3|12(\rho)}||_{KF}),$ $T_i (i=1,2,3)$  denotes the partial transposition with respect to the $ith $ subsystem, $R_{i|jk(\rho)}$ is bipartite realignment operation on subsystem $i$ and subsystem $j,k$, $i\ne j\ne k=1,2,3,$ and $||\cdot||_{KF}$ denotes the trace norm of a matrix.}
 
\textbf{Proof.}
Based on the result of the lower bound for the bipartite concurrence  in \cite{chen kai},
$$C_{A|B}(\rho)\geq \sqrt{\frac{2}{d(d-1)}}\left( \max\left\lbrace ||\rho^{T_A}||_{KF},||R(\rho)||_{KF}\right\rbrace -1\right)  ,$$
where $\rho^{T_A}$ stands for a partial transpose with respect to the subsystem $A$ and $R(\rho)$ denotes realignment operation, we can get 
\begin{align}
	&\frac{C_{1|23}(\rho)+C_{2|13}(\rho)+C_{3|12}(\rho)}{\sqrt{2}}
	-2{\sqrt{\frac{d-1}{d}}}\notag\\
	\ge&\sqrt{\frac{1}{d(d-1)}}(||\rho^{T_1}||_{KF}+||\rho^{T_2}||_{KF}+||\rho^{T_3}||_{KF})-3\sqrt{\frac{1}{d(d-1)}}-2{\sqrt{\frac{d-1}{d}}}\notag\\
	=&3\sqrt{\frac{1}{d(d-1)}}\left( M(\rho)-\frac{1+2d}{3}\right) ,\label{ineq1}
\end{align} 
and
\begin{align}
	&\frac{C_{1|23}(\rho)+C_{2|13}(\rho)+C_{3|12}(\rho)}{\sqrt{2}}
	-2{\sqrt{\frac{d-1}{d}}}\notag\\
	\ge&\sqrt{\frac{1}{d(d-1)}}\left( ||R_{1|23(\rho)}||_{KF}+||R_{2|13(\rho)}||_{KF}+||R_{3|12(\rho)}||_{KF}\right) -3\sqrt{\frac{1}{d(d-1)}}-2{\sqrt{\frac{d-1}{d}}}\notag\\
	=&3\sqrt{\frac{1}{d(d-1)}}\left( N(\rho)-\frac{1+2d}{3}\right) .\label{ineq2}
\end{align} 
Combining the inequalities (\ref{ineq1}) and (\ref{ineq2}), we obtain
$$\frac{C_{1|23}(\rho)+C_{2|13}(\rho)+C_{3|12}(\rho)}{\sqrt{2}}
-2{\sqrt{\frac{d-1}{d}}}
\ge L_1(\rho).$$
This proves the superiority of the lower bound for the GME concurrence comparing with than in \cite{Li Ming2}.

\begin{figure}[!htp]
	\centering
	\includegraphics[width=0.7\linewidth]{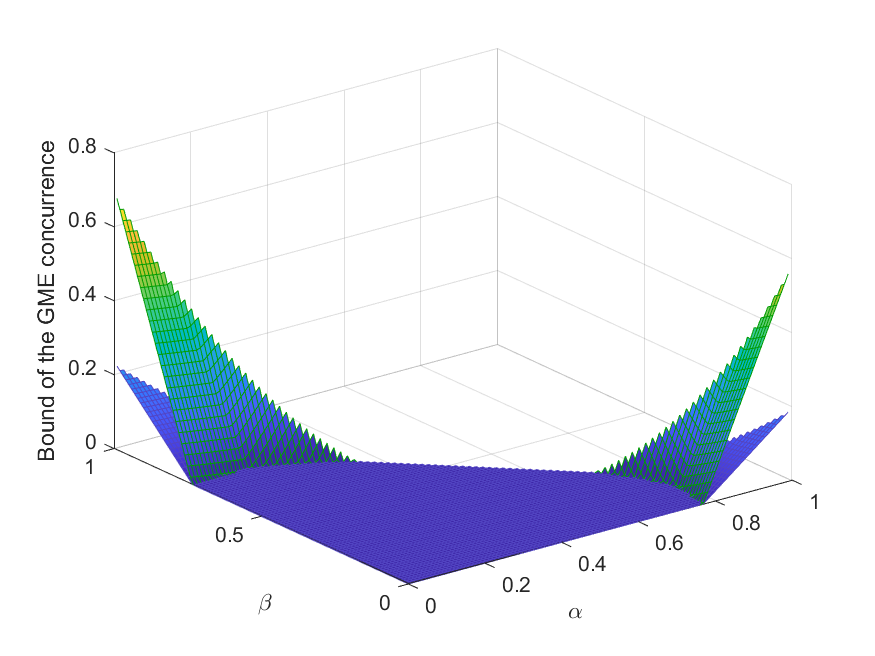}
	\caption{{\small The lower bound of GME concurrence for state $\rho$. The bounds from top to bottom in the figure are given by the lower bounds on the GME concurrence in Theorem 1 (green) and \cite{Li Ming2} (blue), respectively. }}
	\label{fig:1}
\end{figure}

\textbf{Example 1.} Let us consider the mixture of the three-qubit GHZ and W states, $$\rho=\frac{(1-\alpha-\beta)}{8}I+\alpha|GHZ\rangle\langle GHZ|+\beta|W\rangle\langle W|,$$
where $|GHZ\rangle=\frac{1}{\sqrt{2}}(|000\rangle+|111\rangle)$ and $|W\rangle=\frac{1}{\sqrt{3}}(|001\rangle+|010\rangle+|100\rangle).$
For convenience, we estimate the bipartite concurrence in Theorem 1 by the lower bound in \cite{chen kai}. As shown in Figure  \ref{fig:1}, we can see that the lower bound of the GME concurrence  proposed in Theorem 1 is equal to or better than the lower bound in \cite{Li Ming2}.

We then consider the special case of $\rho$ in Example 1 by setting $\alpha=0$,or $\beta=0$. 

\begin{figure}[!h]
	\centering
	\subfigure[]{
		\includegraphics[width=0.47\linewidth]{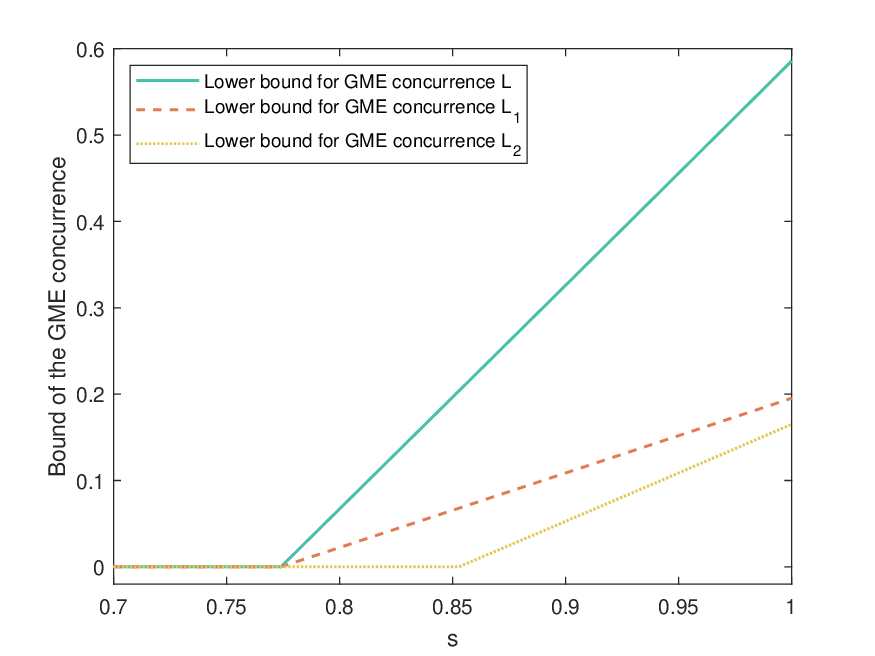}
		\label{fig:subfig1}
	}
	\subfigure[]{
		\includegraphics[width=0.47\linewidth]{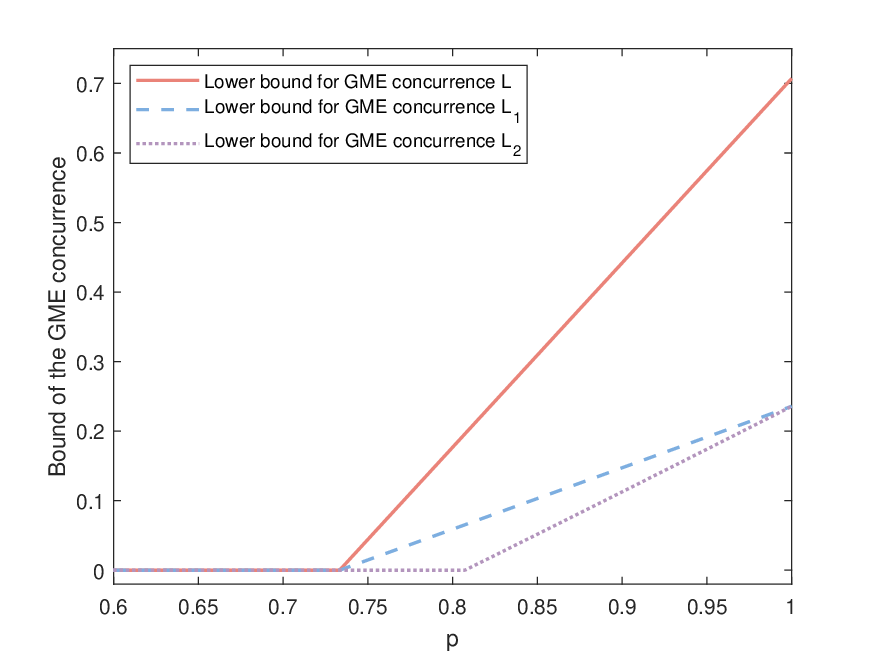}
		\label{fig:subfig2}
	}
	\caption{{\small The comparison of lower bounds of GME concurrence for $ \rho_s$ and $ \rho_p$. In Figure (a), the green solid line denotes the lower bound on the GME concurrence in Theorem 1 for $ \rho_s$, the orange dashed line is the lower bound on the GME concurrence $L_1(\rho)$ in \cite{Li Ming2} for $ \rho_s$, and the yellow dotted line is the lower bound on the GME concurrence $L_2(\rho)$ in \cite{X. Qi}  for $ \rho_s$ when $a=0$ and $b=1$. In Figure (b), the red solid line denotes the lower bounds of GME concurrence in Theorem 1 for $ \rho_p$, the blue dashed line denotes the lower bounds of GME concurrence $L_1(\rho)$ in \cite{Li Ming2} for $ \rho_p$ and the purple dotted line is the lower bounds of GME concurrence $L_2(\rho)$ in  \cite{X. Qi} when $a=b=1$  for $ \rho_p$}.}
	\label{fig:2}
\end{figure}
\textbf{Example 2.} Consider a $3$-qubit W state mixed white noise quantum state and a $3$-qubit GHZ state mixed white noise quantum state,
$$ \rho_s=\frac{1-s}{8}I_8+s|W\rangle\langle W|,$$
$$ \rho_p=\frac{1-p}{8}I_8+p|GHZ\rangle\langle GHZ|, $$
where $0\le s,p\le 1.$ 
We denote the lower bound of the GME concurrence in Theorem 1 as $L$, i.e.,
$$C_{GME}(\rho)\geq \frac{C_{1|23}(\rho)+C_{2|13}(\rho)+C_{3|12}(\rho)}{\sqrt{2}}
-2{\sqrt{\frac{d-1}{d}}}:=L(\rho), $$  and the lower bound of the GME concurrence in \cite{Li Ming2} as $L_1$, i.e.,
$$	C_{GME}(\rho)\geq \frac{1}{\sqrt{d(d-1)}}\left( \max\left\lbrace M(\rho),N(\rho)\right\rbrace -\frac{1+2d}{3}\right):=L_1(\rho). $$ 
From Figure \ref{fig:2} we can clearly see that the lower bound given in Theorem 1 is better to that given in \cite{Li Ming2}. Obviously, our result is also better than Theorem 3 presented in \cite{X. Qi} for
the lower bound of the GME concurrence
$$L_2(\rho)=\frac{1}{\sqrt{d(d-1)}}\left(\mathcal{M} _{a,b }(\rho ) -\sqrt{(1+a^2)(1+b^2)}-\frac{2(d-1)}{3}\right) $$ 
when $a,b$ take on some specific values,
where $a,b\in \mathbb{R},$ $$\mathcal{M} _{a,b }(\rho )=\frac{1}{3} \left ( \left \| \mathcal{M} _{a,b }^{1|23}(\rho ) \right \| + \left \| \mathcal{M} _{a,b }^{2|13}(\rho ) \right \|+\left \| \mathcal{M} _{a,b }^{3|12}(\rho ) \right \|  \right ), $$  $$\mathcal{M} _{a,b }^{A|B}(\rho_{AB} ) =\begin{pmatrix}
	ab & a vec(\rho_B)^T\\
	b vec(\rho_A)& \mathcal{R} (\rho_{AB} )
\end{pmatrix}$$ and $vec(\cdot)$ and $\mathcal{R} (\cdot)$ denote the vectorization of matrix and the realignment of matrix, respectively.

\section{A relationship between multipartite entanglement concurrence and GME concurrence}
In this section, we also first discuss the relationship between the GME concurrence and the multipartite entanglement concurrence in tripartite quantum systems, and generalize it to multipartite systems. This relation gives a lower bound on  $C_{GME}(\rho)$.\\
\par 

\textbf{Theorem 4:}
\emph{Let $\rho  \in {H_{123}} = H_1^d \otimes H_2^d \otimes H_3^d$ be a tripartite qudits quantum state.
Then one has
\be\label{THM5}
C_{GME}(\rho)\geq C_3(\rho)
-{\sqrt{\frac{2(d-1)}{d}}},
\ee
where $C_3(\rho)$ is the tripartite concurrence defined in (\ref{xxxx}).}\\
\textbf{Proof.} We consider pure state first.
For $\rho=|\psi\ra\la\psi|\in H_1^d \otimes H_2^d \otimes H_3^d$, we have
\begin{align*}
	&d(d-1)(1-{\rm tr}\rho_1^2)-d(d-1)\sum_{k=1}^3(1-{\rm tr}\rho_k^2)+2(d-1)^2\\
	=&-d(d-1)\left( (1-{\rm tr}\rho_2^2)+(1-{\rm tr}\rho_3^2)\right) +2(d-1)^2\\
	\geq&0,
\end{align*}
where we have used the inequality $1-{\rm tr}\rho_{k}^2\leq 1-\frac{1}{d}$ for $k=2$ or $3$ to obtain the inequality.
Thus we get
\be 1-{\rm tr}\rho_1^2\geq \sum_{k=1}^3(1-{\rm tr}\rho_k^2)-2\frac{(d-1)}{d}.
\ee
When $\sum_{k=1}^3(1-{\rm tr}\rho_k^2)\geq 2\frac{(d-1)}{d}$, one obtains
\be \sqrt{1-{\rm tr}\rho_1^2}\geq \sqrt{\sum_{k=1}^3(1-{\rm tr}\rho_k^2)}-\sqrt{\frac{2(d-1)}{d}}=C_3(\rho)
-{\sqrt{\frac{2(d-1)}{d}}}.
\ee
Similarly, we get
\be \sqrt{1-{\rm tr}\rho_j^2}\geq C_3(\rho)-\sqrt{\frac{2(d-1)}{d}},
\ee
for $j=2,3.$

Then according to the definition of GME concurrence, we derive
\be
C_{GME}(|\psi\ra)\geq C_3(\rho)-\sqrt{\frac{2(d-1)}{d}}.
\ee

Now consider a mixed state $\rho\in H_1^d \otimes H_2^d \otimes H_3^d$ with the optimal ensemble decomposition
$\rho=\sum_{\alpha}p_{\alpha}|\psi_{\alpha}\ra\la\psi_{\alpha}|$, which minimizes the GME concurrence.
Using the convexity of concurrence, one obtains
\begin{align*}
	&C_{GME}(\rho)=\sum_{\{p_{\alpha},|\psi_{\alpha}\ra\}}p_{\alpha}C_{GME}\left( |\psi_{\alpha}\ra\right) \notag\\
	&\geq \sum_{\alpha}p_{\alpha}\left( C_3(|\psi_{\alpha}\ra)-\sqrt{\frac{2(d-1)}{d}}\right) \notag\\
	&\geq C_3(\rho)-\sqrt{\frac{2(d-1)}{d}},
\end{align*}
which proofs the result of the Theorem.
\hfill \rule{1ex}{1ex}
\par 
The above Theorem gives a relation between the genuine entanglement concurrence  and the entanglement concurrence for tripartite quantum states. Next, we generalize the relation between GME concurrence and multipartite entanglement concurrence to give a lower bound on GME concurrence.\\
\par

\textbf{Theorem 5:}
\emph{Let $\rho  \in {H_{12\cdots N}} = H_1^d \otimes H_2^d \otimes\cdots\otimes H_N^d$ be a $N$-partite qudits quantum state.
Then one has
\be\label{THM4}
C_{GME}(\rho)\geq 2^{\frac{N-3}{2}}C_{N}(\rho)
-{\sqrt{\frac{(2^{N-1}-2)(d-1)}{d}}},
\ee
where $C_{N}(\rho)$ is the $N$-partite concurrence defined in (\ref{xxxx}). } \\
\textbf{Proof.}
We consider the pure state first.
For $\rho=|\psi\ra\la\psi|\in {H_{12\cdots N}} = H_1^d \otimes H_2^d \otimes\cdots\otimes H_N^d$ and for all $2^{N-1}-1 $ different partitions $x|\overline{x}$, $x,\overline{x}\in \mathcal{P}(\mathcal{S})\setminus \left\lbrace \varnothing,\mathcal{S}\right\rbrace $, we have
\begin{align*}
	&d(d-1)(1-{\rm tr}\rho_1^2)-d(d-1)\sum_{x}(1-{\rm tr}\rho_x^2)+(2^{N-1}-2)(d-1)^2\\
	=&-d(d-1)\sum_{x\ne 1}(1-{\rm tr}\rho_x^2)+(2^{N-1}-2)(d-1)^2\\
	\geq&0,
\end{align*}
where we have used the inequality $1-{\rm tr}\rho_{k}^2\leq 1-\frac{1}{d}$ for all the different partitions $x|\overline{x}$, $x,\overline{x}\in \mathcal{P}(\mathcal{S})\setminus \left\lbrace \varnothing,\mathcal{S},\left\lbrace 1\right\rbrace \right\rbrace $ to obtain the inequality.
Thus we get
\be 1-{\rm tr}\rho_1^2\geq \sum_{x}(1-{\rm tr}\rho_x^2)-(2^{N-1}-2)\frac{(d-1)}{d}.\label{tr}
\ee
When $\sum_{x}(1-{\rm tr}\rho_x^2)\geq (2^{N-1}-2)\frac{(d-1)}{d}$, one obtains
\be \sqrt{1-{\rm tr}\rho_1^2}\geq \sqrt{\sum_{x}(1-{\rm tr}\rho_x^2)}-\sqrt{(2^{N-1}-2)\frac{(d-1)}{d}}=2^{\frac{N-3}{2}}C_N(\rho)
-{\sqrt{\frac{(2^{N-1}-2)(d-1)}{d}}}.
\ee
Similarly, we get
\be \sqrt{1-{\rm tr}\rho_j^2}\geq 2^{\frac{N-3}{2}}C_N(\rho)-\sqrt{\frac{(2^{N-1}-2)(d-1)}{d}},
\ee
for different partitions $j|\overline{j}$, $j,\overline{j}\in \mathcal{P}(\mathcal{S})\setminus \left\lbrace \varnothing,\mathcal{S},\left\lbrace 1\right\rbrace \right\rbrace.$

Then according to the definition of GME concurrence, we obtain
\be
C_{GME}(|\psi\ra)\geq 2^{\frac{N-3}{2}}C_N(\rho)-\sqrt{\frac{(2^{N-1}-2)(d-1)}{d}}.
\ee

Finally we consider a mixed state $\rho\in H_1^d \otimes H_2^d \otimes\cdots\otimes H_N^d$ that satisfies the optimal decomposition
$\rho=\sum_{\alpha}p_{\alpha}|\psi_{\alpha}\ra\la\psi_{\alpha}|$ such that the GME concurrence is minimized.
Through the convexity property of concurrence, one gets that
\begin{eqnarray*}
	&&C_{GME}(\rho)=\sum_{\{p_{\alpha},|\psi_{\alpha}\ra\}}p_{\alpha}C_{GME}(|\psi_{\alpha}\ra)\\
	&&\geq \sum_{\alpha}p_{\alpha}\left( 2^{\frac{N-3}{2}}C_N(|\psi_{\alpha}\ra)-\sqrt{\frac{(2^{N-1}-2)(d-1)}{d}}\right) \\
	&&\geq 2^{\frac{N-3}{2}}C_N(\rho)-\sqrt{\frac{(2^{N-1}-2)(d-1)}{d}},
\end{eqnarray*}
which gives us the result of the Theorem.
\hfill \rule{1ex}{1ex}

\textbf{Remark 2.} When $N=3$, Theorem 5 reproduces as Theorem 4, which gives the relationship between the GME concurrence and the multipartite entanglement concurrence.\par 


\section{Applications}\label{sec3}

Based on the results of Section \Rmnum{3} and \Rmnum{4}, we may consider the relationship between other entanglement measures and the GME concurrence. We first consider the multipartite tangle \cite{P. Rungta,J. Wang1} that is closely related with concurrence. For a multipartite pure quantum state $|\psi\rangle\langle \psi|\in H_1\otimes H_2 \otimes\cdots\otimes H_N$, multipartite tangle is represented by the square of multipartite concurrence, i.e.,
\begin{align}
	\tau_N(|\psi\rangle\langle \psi|)=C_N^2(|\psi\rangle\langle \psi|)=2^{2-N}\left( \sum_{\alpha}(1-{\rm tr}[\rho_{\alpha}^{2}]\right) ,\label{tau}
\end{align}
where $\rho_{\alpha}$ denotes all the reduced matrices. For mixed states, we can obtain the relation between the multipartite tangle and multipartite concurrence as 
$$\tau_N(\rho )=\min_{\{p_i,|\varphi _i\rangle \}}\sum_{i}p_i\tau_N(|\varphi _i\rangle\langle \varphi _i| )= \min_{\{p_i,|\varphi _i\rangle \}}\sum_{i}p_i C_N^2(|\varphi _i\rangle\langle \varphi _i|),$$
where the minimum is taken over all possible ensemble decompositions of $\rho  = \sum\limits_i {{p_i}\left| {{\varphi _i}} \right\rangle } \left\langle {{\varphi _i}} \right|$, ${p_i} \ge 0$ and $\sum\limits_i {{p_i}}  = 1$. In the following we will give the relation between  multipartite tangle and GME concurrence.\\ \par 
\textbf{Theorem 6:}\emph{ Let the mixed state $\rho\in  H_1^d \otimes H_2^d \otimes\cdots\otimes H_N^d$ be an $N$-partite qudits state, we can  get 
\begin{align*}
	C_{GME}(\rho)\geq 2^{N-3}\tau_N(\rho)-\frac{(2^{N-1}-2)(d-1)}{d}.
\end{align*}}
\textbf{Proof.}
We consider the pure state first.
For $\rho=|\psi\ra\la\psi|\in {H_{12\cdots N}} = H_1^d \otimes H_2^d \otimes\cdots\otimes H_N^d,$ according to (\ref{tr}), 
\be 1-{\rm tr}\rho_1^2\geq \sum_{x}(1-{\rm tr}\rho_x^2)-(2^{N-1}-2)\frac{(d-1)}{d},
\ee
 and  from (\ref{tau}), one obtains
\be 1-{\rm tr}\rho_1^2\geq \sum_{k}(1-{\rm tr}\rho_k^2)-(2^{N-1}-2)\frac{(d-1)}{d}=2^{N-3}\tau_N(\rho)
-\frac{(2^{N-1}-2)(d-1)}{d}.
\ee
Similarly, we get
\be 1-{\rm tr}\rho_j^2\geq 2^{N-3}\tau_N(\rho)
-\frac{(2^{N-1}-2)(d-1)}{d},
\ee
for different partitions $j|\overline{j}$, $j,\overline{j}\in \mathcal{P}(\mathcal{S})\setminus \left\lbrace \varnothing,\mathcal{S},\left\lbrace 1\right\rbrace \right\rbrace.$

Then according to the definition of GME concurrence, we have
\be
C_{GME}(|\psi\ra)\geq 2^{N-3}\tau_N(\rho)
-\frac{(2^{N-1}-2)(d-1)}{d}.
\ee

Finally we consider a mixed state $\rho\in H_1^d \otimes H_2^d \otimes\cdots\otimes H_N^d$ with the optimal ensemble decomposition
$\rho=\sum_{\alpha}p_{\alpha}|\psi_{\alpha}\ra\la\psi_{\alpha}|$ such that the GME concurrence attains its minimum.
 Using the convexity property of concurrence, one has that
\begin{eqnarray*}
	&&C_{GME}(\rho)=\sum_{\{p_{\alpha},|\psi_{\alpha}\ra\}}p_{\alpha}C_{GME}(|\psi_{\alpha}\ra)\\
	&&\geq \sum_{\alpha}p_{\alpha}\left( 2^{N-3}\tau_N(|\psi_{\alpha}\ra)
	-\frac{(2^{N-1}-2)(d-1)}{d}\right) \\
	&&\geq 2^{N-3}\tau_N(\rho)
	-\frac{(2^{N-1}-2)(d-1)}{d},
\end{eqnarray*}
which completes the proof of the Theorem.
\hfill \rule{1ex}{1ex}\\ \par
Next we consider global negativity \cite{S.S. Sharma,L. Zhang} associated with the bipartite concurrence. For any $N$-qubit pure state $\rho_{12\cdots N}=|\psi\rangle\langle\psi|$, the global negativity
is defined as $$N^p(|\psi\rangle)=\sqrt{2(1-{\rm tr}\rho_p^2)}=C_{p|\overline{p}}(|\psi\rangle),$$
where $p,\overline{p}\in\left\lbrace 1,2,\cdots,N\right\rbrace ,p=\left\lbrace 1,2,\cdots,N\right\rbrace \setminus \left\lbrace \overline p \right\rbrace .$ Below we present the relationship between GME concurrence and global negativity.\\
\par 
\textbf{Theorem 7:}\emph{ Let the state $\rho\in  H_1^d \otimes H_2^d \otimes\cdots\otimes H_N^d$, we get 
\begin{align*}
	C_{GME}(\rho)\geq \frac{1}{\sqrt{2}}\sum_{x}N^x(\rho)-(2^{N-1}-2)\sqrt{\frac{d-1}{d}}.
\end{align*}}\\
\textbf{Proof.}
We consider the pure state first.
For $\rho=|\psi\ra\la\psi|\in {H_{12\cdots N}} = H_1^d \otimes H_2^d \otimes\cdots\otimes H_N^d.$ According to (\ref{11}), we have
\begin{eqnarray*}
	&&\sqrt{1-{\rm tr}\rho_k^2}\geq \frac{1}{\sqrt{2}}\sum_{x}C_{x|\overline{x}}(|\psi\rangle)-(2^{N-1}-2)\sqrt{\frac{d-1}{d}}\\
	&&=\frac{1}{\sqrt{2}}\sum_{x}N^x(|\psi\rangle)-(2^{N-1}-2)\sqrt{\frac{d-1}{d}},
\end{eqnarray*}
where the summation term is taken over all the different bipartite partitions $k|\overline{k}$. Then according to the definition of GME concurrence, we obtain 
\be
C_{GME}(|\psi\rangle)\geq \frac{1}{\sqrt{2}}\sum_{x}N^x(|\psi\rangle)-(2^{N-1}-2)\sqrt{\frac{d-1}{d}}.
\ee
Now consider a mixed state $\rho\in  H_1^d \otimes H_2^d \otimes\cdots\otimes H_N^d$ with the optimal ensemble decomposition
$\rho=\sum_{\alpha}p_{\alpha}|\psi_{\alpha}\ra\la\psi_{\alpha}|$ such that the GME concurrence attains its minimum.
By using the convexity property of concurrence, we get 
\begin{align*}
	&C_{GME}(\rho)=\sum_{\{p_{\alpha},|\psi_{\alpha}\ra\}}p_{\alpha}C_{GME}(|\psi_{\alpha}\ra)\\
	&\geq \sum_{\alpha}p_{\alpha}\left( \frac{1}{\sqrt{2}}\sum_{x}N^x(|\psi_{\alpha}\rangle)-(2^{N-1}-2)\sqrt{\frac{d-1}{d}}
	\right) \\
	&\geq\frac{1}{\sqrt{2}}\sum_{x}N^x(\rho)-(2^{N-1}-2)\sqrt{\frac{d-1}{d}},
\end{align*}
which leads to the result of the Theorem.
\hfill \rule{1ex}{1ex}\par

\section{Conclusions and Discussions}\label{sec4}

While the GME offers significant advantage in quantum information processing tasks, it is a basic and fundamental problem in quantum theory to detect and measure such physical resource. 
In this paper we investigate the interconnection between GME concurrence and bipartite entanglement concurrence, and between GME concurrence and multipartite entanglement concurrence, which give lower bounds on GME concurrence, respectively. Among them we prove that the introduced lower bound for the GME concurrence is better than some existing results. Finally we explore the relationship between the GME concurrence and other related entanglement measures, which give lower bounds for the GME concurrence. These will be useful for exploring entanglement measures and studying GME in the future.

\bigskip
\noindent{\bf Acknowledgments}\, \, This work is supported by NSFC No.11701568, 12075159, the Fundamental Research Funds for the Central
Universities No.22CX03005A, 24CX03003A, the Shandong Provincial Natural Science Foundation for Quantum Science No. ZR2021LLZ002, Beijing Natural Science Foundation (Z190005), the Academician Innovation Platform of Hainan Province, and Academy for Multidisciplinary Studies, Capital Normal University.

\end{document}